\documentclass[pra,aps,twocolumn,amsmath,amssymb]{revtex4}
\usepackage{graphicx}
\pdfoutput=1
\usepackage{amsmath}
\usepackage{color}
\usepackage{float}
\setcitestyle{super}
\usepackage[english]{babel}
\usepackage{color,soul}

\begin{document}
\title{Mixed type I and type II superconductivity due to intrinsic electronic inhomogeneities in the type II Dirac semimetal PdTe$_2$ }

\author{Anshu Sirohi}
\email{anshusirohi@iisermohali.ac.in}

\author{Shekhar Das, Rajeswari Roy Chowdhuri, Amit, Yogesh Singh, Sirshendu Gayen}

\author{Goutam Sheet}
\email{goutam@iisermohali.ac.in}

\affiliation{Department of Physical Sciences, Indian Institute of Science Education and Research(IISER) Mohali, Sector 81, S. A. S. Nagar, Manauli, PO: 140306, India.}

\begin{abstract}

\textbf{The type II Dirac semimetal PdTe$_2$ is unique in the family of topological parent materials because it displays a superconducting ground state below 1.7 K. Despite wide speculations on the possibility of an unconventional topological superconducting phase, tunneling and heat capacity measurements revealed that the superconducting phase of PdTe$_2$ follows predictions of the microscopic theory of Bardeen, Cooper and Shriefer (BCS) for conventional superconductors. The superconducting phase in PdTe$_2$ is further interesting because it also displays properties that are characteristics of type-I superconductors and are generally unexpected for binary compounds. Here, from scanning tunneling spectroscopic measurements we show that the surface of PdTe$_2$ displays intrinsic electronic inhomegenities in the normal state which leads to a mixed type I and type II superconducting behaviour along with a spatial distribution of critical fields in the superconducting state. Understanding of the origin of such inhomogeneities may be important for understanding the topological properties of PdTe$_2$ in the normal state.}

\end{abstract}

\maketitle
PdTe$_2$ has been known to be a superconductor for almost six decades.\cite{JG, Kjeshus, Raub, Roberts} But, due to a very low critical temperature ($\sim$ 1.7 K), the details of the superconducting phase of PdTe$_2$ did not receive much attention. The superconducting properties of PdTe$_2$ attracted renewed attention of the community following recent discovery of complex topological features in the band structure of PdTe$_2$ in it's non-superconducting normal state\cite{Noh, Fei, LYan}. This discovery naturally led to the question of the possibility of a topological character of the superconducting phase of PdTe$_2$.\cite{Fei} This question is extremely important because a single material showing both topological character and superconductivity in it's parental stoichiometric phase is thought to be the best candidate to show topological superconductivity\cite{Ludwig, Zhang1, Zhang2, Kane, YAndo, MSato, SDS1, SDS2, SDS3, SDS4, Leo1}. Immediately after the discovery of the topological nature of PdTe$_2$,\cite{Noh, Fei} transport and magnetization experiments revealed a peculiar type-I like superconducting phase but with multiple critical fields ($H_c$).\cite{HLeng, P2} The possible type-I nature of superconductivity in PdTe$_2$ is unique because only elemental metals are known to show type-I superconductivity and superconductivity in binary compounds and alloys are generally seen to be of type II with perhaps only one exception of TaSi$_2$\cite{TaSi2}. The observation of multiple critical fields in PdTe$_2$ was attributed to the possibility of an anisotropic superconducting order parameter and to an enhanced critical field of the surface sheath ($H_c^s$) of PdTe$_2$.\cite{HLeng, P2} However, the measured $H_c^s$ turned out to be around 3000 G which is one order of magnitude higher than $H_c \sim$ 250 G. This deviates from the expected value for a standard Saint-James-deGennes (SJdG) surface critical field which, in principle, should be less than 1.7 $H_c$.\cite{DSJ1, DSJ2, Tinkham} This mismatch of the measured $H_c^s$ with standard theory was adduced to the possible unconventional topological nature of the superconducting phase.\cite{HLeng, P2} More recently, scanning tunneling spectroscopy (STS) experiments revealed a BCS-like\cite{BCS} conventional superconducting gap in PdTe$_2$\cite{Sdas} which was further confirmed by subsequent STM experiments by other groups\cite{Clark}, penetration depth measurements\cite{P1, P2} and heat capacity measurements\cite{Amit}. However, in the STS experiments it was also seen that the upper critical field has a distribution on a pristine surface of the PdTe$_2$ crystals. The range of values of $H_c$ that were measured in magnetic field dependent STS experiments varied from 220 G to 4 Tesla. The  probability of getting spectra with a small $H_c$ of $\sim$ 250 G was maximum and was significantly higher than that with a large $H_c$. Since STM is a surface sensitive experiment, the observation of such a distribution confirms that all the smaller $H_c$ and larger $H_c$ appear on the pristine surface itself. Therefore, the idea that the higher value of $H_c$ could be due to an enhanced sheath critical field can be ruled out and the origin of multiple $H_c$ in PdTe$_2$ should be investigated in further detail.

\begin{figure}[h!]
		\includegraphics[width=0.5\textwidth]{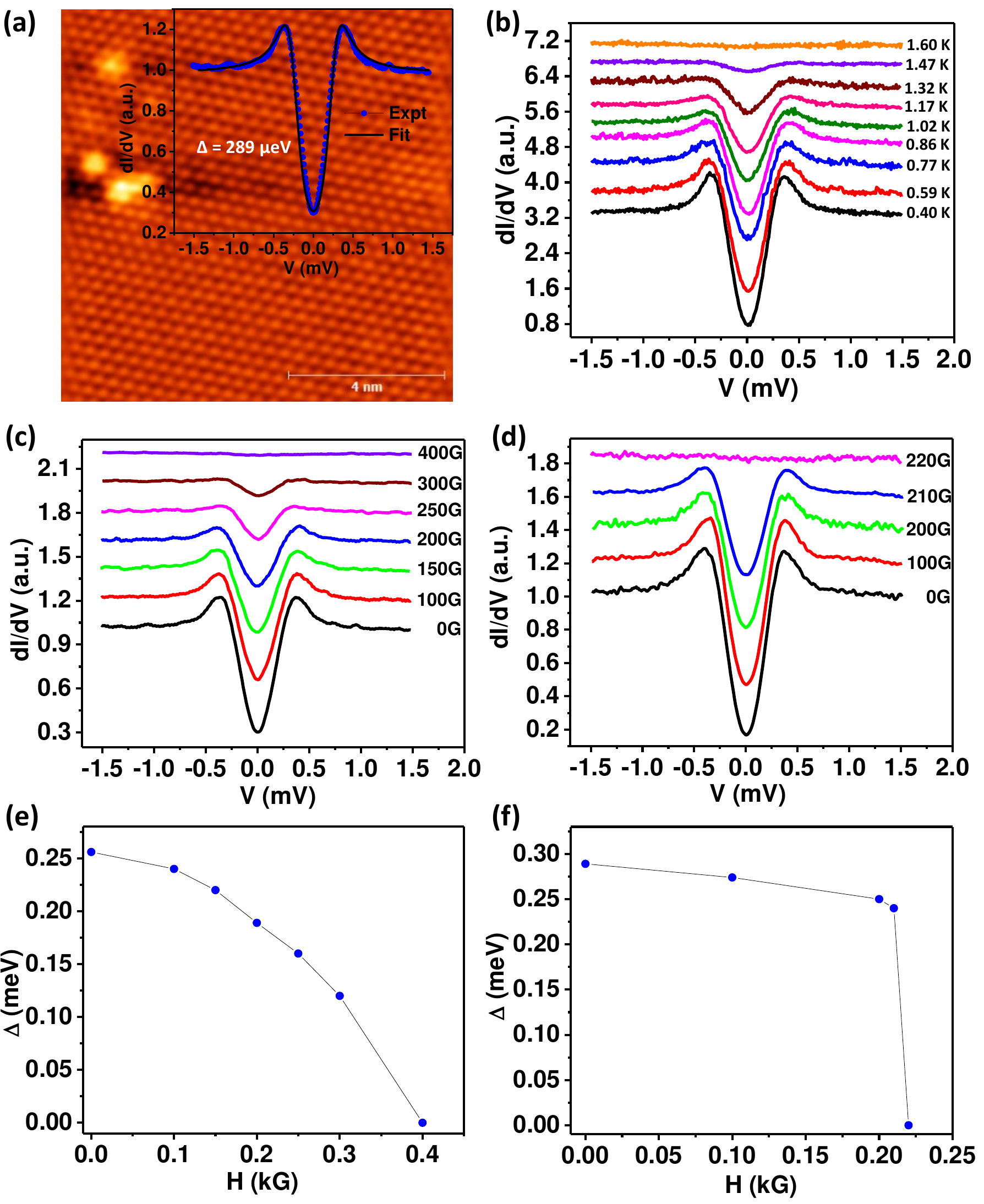}
		\caption{a) The atomic resolution STM image of (111) cleaved PdTe$_2$ with surface defects (see the bright spots) at T = 385 mK. $inset:$ A representative tunneling conductance curve with fit using Dyne's formula showing superconducting energy gap of 289 $\mu$eV. (b) Temperature dependence of the spectrum (as in $inset$ of (a). (c) Magnetic field dependence of a spectrum showing type II behaviour with $H_c$ = 400 G, and (d) Magnetic field dependence of a spectrum showing type I behaviour with $H_c$ = 220 G. The spectra at different magnetic fields in (b, c, d) have been vertically shifted for visual clarity.}	
	\label{Figure 4}
\end{figure}

In this paper, we present a detailed study of the magnetic field dependence of the superconducting energy gap at multiple points on the surface of high quality single crystals of PdTe$_2$ by STS experiments. As mentioned before, our studies presented here also show that there is a distribution of critical fields in PdTe$_2$ and the critical field varies over a broad range starting from as low as 220 G to a high value of 4 Tesla. The points on the crystals surface showing low critical field reveal a type I behaviour while the points showing high critical fields show a gradual disappearance of the superconducting features with increasing magnetic fields as in type II superconductors. We also observe clear signature of intrinsic electronic inhomogeneities in the conductance maps recorded in the normal state of PdTe$_2$. The distribution of superconducting properties in the superconducting state can be attributed to such inhomogeneities. This is remarkably similar to the earlier observations made on the high $T_c$ cuprate superconductor BSCCO.\cite{SDavis}

The experiments were carried out in an ultra-high-vacuum (UHV) Scanning Tunnelling Microscope (STM) equipped with low temperature sample cleaving facility and the STM works down to 380 mK. All the STM and STS measurements were carried out on high quality single crystals of PdTe$_2$  which were cleaved at 77K by the $in-situ$ cleaver. The unit cell of PdTe$_2$ is hexagonal with one Pd and two Te atoms as in CdI$_2$ (space group $P\overline{3}m1$).\cite{LT} The hexagonal structure of PdTe$_2$ is clearly seen in the atomically resolved images captured at 385 mK at 400 mV and 250 pA as shown in Figure 1(a). Defects on the surface of the crystal have also been clearly resolved. 

\begin{figure}[h!]
		\includegraphics[width=0.45\textwidth]{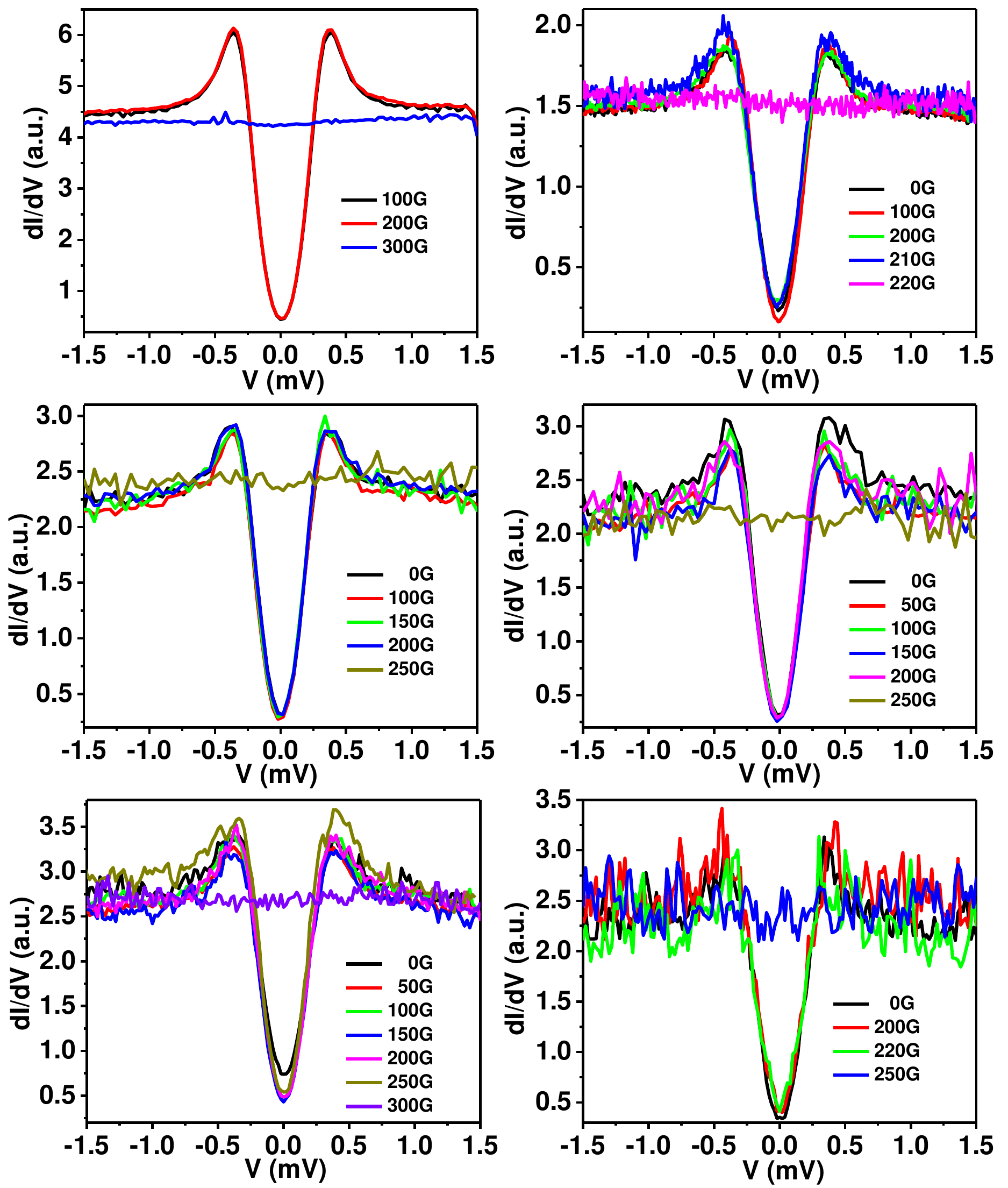}
		\caption{ Six representative spectra with low critical magnetic fields and their evolution with magnetic fields. Such spectra also show a distribution ranging between 220-300G.}	
	\label{Figure 4}
\end{figure}

In the superconducting state, the STS spectrum shows smooth variation of BCS-like density of states with energy ($inset$ of Figure 1(a)) along with two coherence peaks symmetric about $V$ = 0. The coherence peaks appear near the superconducting energy gap ($\Delta$) in STS spectra on a superconductor. The exact amplitude of the superconducting energy gap is determined by fitting the tunneling conductance ($dI/dV$) vs. $V$ curves using  Dyne's formula which gives density of states (DOS) $N_s(E) = Re\left(\frac{(E-i\Gamma)}{\sqrt{(E-i\Gamma)^2-\Delta^2}}\right)$, where $\Gamma$ is an effective broadening parameter incorporated to take care of slight broadening of the BCS density of states possibly due to finite quasi-particle life time.\cite{Dynes} Following this procedure, we have measured a superconducting energy gap of $\sim$289 $\mu eV$ which is slightly lower than the previously reported results\cite{Sdas}. As we will show later, this variation of the superconducting energy gap on PdTe$_2$ could also be a consequence of intrinsic electronic inhomogeneities. Though the measured $\Delta$ is slightly lower, the temperature dependence of the spectra shows that the gap closes smoothly with increasing temperature as per BCS theory (Figure 1(c)).

\begin{figure}[h!]
	\centering
		\includegraphics[width=0.5\textwidth]{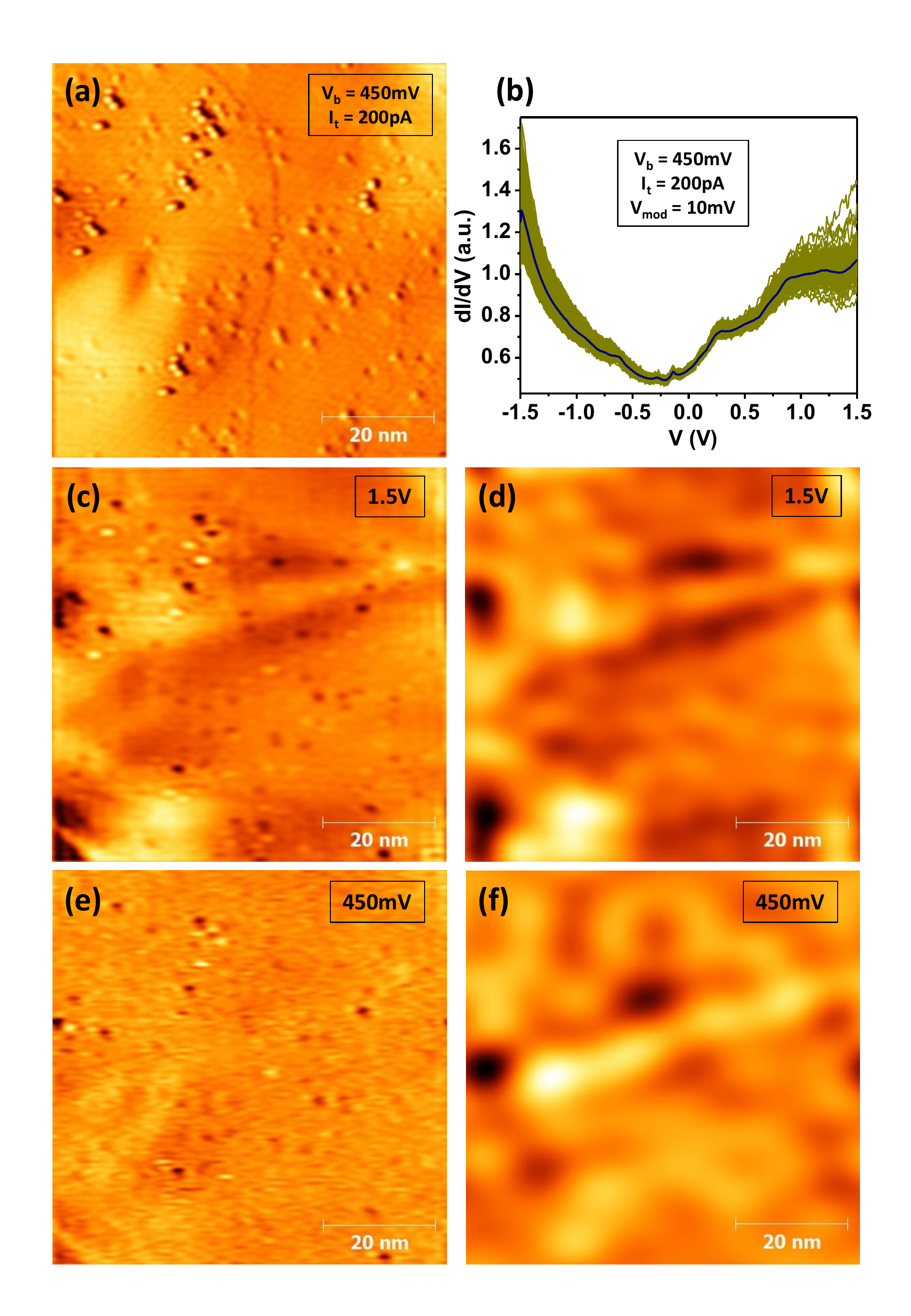}
		\caption{(a) Large area topographic image of cleaved surface of PdTe$_2$ crystal measured in constant-current ($I$ = 200 pA) mode (70nm x 70nm) at 3.5 K. (b) Some representative conductance spectra over the same area as shown in (a). (c), (e) Conductance maps at 1.5V and 450mV respectively shows LDOS at the same area as shown in (a). (d), (e) The variation of the Fourier filtered LDOS reflecting the intrinsic electronic inhomogeneities in PdTe$_2$.}	
	\label{Figure 4}
\end{figure}

As it was discussed before, a spatial distribution of critical field ($H_c$) is observed on the surface of PdTe$_2$\cite{Sdas} where the points showing lower critical field also showed type-I superconductivity. In order to understand the details of such distribution, we have repeated the spectroscopic measurements on a large number of points and investigated the magnetic field dependence of the spectra. A representative spectrum with relatively higher critical field ($\sim$ 400 G) and the field dependence of the same is shown in Figure 1(c) where superconducting gap is seen to decrease continuously with increasing magnetic field. The smooth evolution is reflected in the magnetic field dependence of $\Delta$ in Figure 1(e). Another representative spectrum is shown in Figure 1(d), where a first order disappearance of the superconducting gap is observed at a critical field of 220 G. Such a sudden disappearance of superconductivity with increasing magnetic field (Figure 1(f)) is due to type I nature of superconductivity at those points. This is statistically observed at large number of points for all of which the critical magnetic field remained low and fell in a range between 220 G to 300 G. In Figure 2 we show a representative set of such spectra where sudden disappearance of the spectral features associated with superconductivity are seen.  On the other hand, for all the spectra of critical field higher than 300 G show type-II behaviour. It should be noted that in the past similar observations of local first order transitions corresponding to type-I superconductivity were reported in point-contact spectroscopic measurements on some of the elemental superconductors.\cite{Naidyuk} Our observations, therefore, indicate that the critical magnetic field has an inhomogeneous distribution on the pristine surface of PdTe$_2$ and the electromagnetic properties of the superconductor also varies from point to point.

A distribution of criticial field ($H_c$) directly hints to a distribution of superconducting coherence length ($\xi$) over the surface of the crystal as per the relationship $\xi^2 =\Phi_0/2\pi H_c $, where $\Phi_0$ is single quantum of magnetic flux. $\xi$, on the other hand, may vary spatially if there is a variation of the mean free path ($l$) over the sample surface, as expected in a superconductor with localized disorders following the relationship $1/\xi =  1/\xi_0 + 1/l$, where $\xi_0$ is the intrinsic coherence lengh at zero temperature as per the description in BCS theory.\cite{Tinkham} From this understanding, it is natural to believe that the distribution of the superconducting properties may be correlated with the distribution of the defects on the surface of PdTe$_2$. However, in our experiments no such correlation was observed. Threfore, it is possible that the distribution arises from a more intrinsic property of the system. An intrinstic electronic inhomogeneity, as in case of some of the cuprates where a prominent electronic inhomogeneity is observed\cite{SDavis}, is the most probable factor that must be considered first.

In order to investigate the possibility of such intrinsic  inhomogeneities in PdTe$_2$, we have investigated the local density of states (LDOS) maps over a large area and studied the variation of the same with respect to the variation  of topography due to different types of defects. In Figure 3(a), we show a large area topograph of the surface of PdTe$_2$ where a substantial number of defects are observed. In Figure 3(b), we show a set of $dI/dV$ vs. $V$ spectra obtained over the same area as in Figure 3(a) at a large number of pixels. In Figure 3(c), we present a conductance map at a particular energy (1.5 eV), where the contribution of the defect states appears as small (circular) bright/dark spots and such spots are directly correlated with the defects imaged in real space (Figure 3(a)). A close inspection of Figure 3(c) reveals that in addition to the variation of the LDOS due to the presence of the topographic defects, additional features are observed in the background. Such a background may originate from inhomogeneities of intrinsic nature. The topographic image was obtained in ``constant current" mode where the tunneling current varies exponentially with the width of the tunneling barrier or the tip-sample distance. The tunneling current is also related to the integrated LDOS. Consequently, the topographic image thus recorded is also expected to contain information of both topographic height distribution and the LDOS. In the topographic image shown in Figure 3(a), we indeed observe additional contrast apart from the modulation due to topographic defects. A similar background is also visible in the LDOS map, where a clear bright/dark contrast in the background of the states corresponding to the surface defects is visible. We used Fourier filtering of the  conductance map and extracted the background signal alone. The maps of the extracted signal (Figure 3(d) and Figure 3(f)) clearly reveal inhomogeneities of the LDOS. A comparison of the maps at different energies (1.5 eV in Figure 3(d) and 450 meV in Figure 3(f)) show that the background inhomogeneity of LDOS also evolves with energy. This inhomogeneity might be because of puddling of electrons and holes as is seen in certain 2-Dimensional systems\cite{Graphene1, Graphene2} and in topological insulators\cite{BSTS1, BSTS2} with Dirac point lying close to the chemical potential.

The intrinsic inhomogeneity mentioned above may lead to a distribution of the superfluid density on the surface of PdTe$_2$ when the system is in it's superconducting state. This is confirmed by the observation of a spatial variation of the superconducting energy gap ($\Delta$). The amplitude of $\Delta$ measured at different points varied over a range between 250 $\mu$eV and 350 $\mu$eV\cite{Sdas}. The inhomogeneity may also lead to the distribution of the coherence length ($\xi$) or the critical field that we have discussed above. The points where the LDOS is low, $\xi$ is expected to be lower leading to a higher critical field and the corresponding type II nature of superconductivity. The points of higher LDOS due to such inhomogeneity leads to the type I behaviour. 

Though the origin of the intrinsic electronic inhomogeneity in PdTe$_2$ is not clear at the moment, one might speculate the inhomogeneity to be associated with phase seperation in real space as it was earlier seen in disordered superconductors both theoretically \cite{AmitGhoshal} and experimentally \cite{Pratap}. It is understood that for such phase seperation, two competeing phenomena must be present in the system. In case of PdTE$_2$, since it is clear that though the normal state shows topologically non-trivial behaviour, the superconducting phase is non-topological in nature, it is possible that the low temperature (below 1.7 K)  superconducting order competes with the topological protection. Detailed theoretical calculations would be necessary to verify the validity of this argument.

In conclusion, from STM and STS experiments we have shown that the surface of the single crystals of PdTe$_2$ host electronic inhomogeneities in the normal state. When the system makes a superconducting transition, the inhomogeneous density of states give rise to a spatially varying superfluid density leading to variation of the coherence lengh $\xi$. The variation of the superfluid density is confirmed by the observation of a distribution of the superconducting energy gap ($\Delta$) on the surface. Since the (upper) critical field is directly related to $\xi$, the distribution also causes a spatial distribution of the critical magnetic fields. For certain values of $\xi$ the superconductivity of PdTe$_2$ at certain points falls in the type I regime. At other places a type II behavior is observed. Therefore, this Letter explains the mixed type I and type II superconducting behaviour that was earlier observed in the type II Dirac semimetal PdTe$_2$. The intrinsic electronic inhomogeneities on PdTe$_2$ that we reported here should be considered theoretically and find out whether this could be related to the topologically non-trivial band structure of PdTe$_2$.

We thank Sanjeev Kumar for fruitful discussions. GS would like to acknowledge financial support from the research grant of Swarnajayanti fellowship awarded by the Department of Science and Technology (DST), Govt. of India under the grant number DST/SJF/PSA-01/2015-16.

\end{document}